# On Schrödinger equations
# with concentrated nonlinearities

Giovanni Jona-Lasinio

*Dipartimento di Fisica, Università di Roma "La Sapienza",*
*Piazzale A. Moro 2, Roma, Italy 00185*
*and Centro Linceo Interdisciplinare, via della Lungara 10, Roma, Italy 00165*

Carlo Presilla

*Dipartimento di Fisica, Università di Roma "La Sapienza",*
*Piazzale A. Moro 2, Roma, Italy 00185*

Johannes Sjöstrand[1]
*Département de Mathématiques, Bât. 425, Université de Paris-Sud,*
*Orsay, France 91405*
*and U.R.A. 760, C.N.R.S.*
(January 10, 1995)

## Abstract

Schrödinger equations with nonlinearities concentrated in some regions of space are good models of various physical situations and have interesting mathematical properties. We show that in the semiclassical limit it is possible to separate the relevant degrees of freedom by noticing that in the regions where the nonlinearities are effective all states are suppressed but the metastable ones (resonances). In this way the description of the nonlinear regions is reduced to ordinary differential equations weakly coupled to standard Schrödinger equations valid in the linear regions. The idea is illustrated through the study of a prototype equation recently proposed for resonant tunneling of electrons through a double barrier heterostructure and for which nonlinear oscillations have been numerically predicted.

03.65.-w, 73.40.Gk

Typeset using REVTEX

# I. INTRODUCTION

In a recent paper [2] a nonlinear Schrödinger equation was proposed to model the effect of the accumulation of electric charge in the resonant tunneling of electrons through a double barrier heterostructure [3]. The main feature of this equation is that the nonlinearity is concentrated only in the region where the resonant state is localised, that is within the two barriers. The equation of Ref. [2] is a prototype of a class of nonlinear Schrödinger equations which can be used to model several physical situations common in mesoscopic physics, e.g. superlattices. In this paper we show that as far as the evolution of the solutions over a finite time interval is concerned, it is possible to carry out an adiabatic theory which simplifies considerably the problem. In fact in the semiclassical limit it is possible to separate the relevant degrees of freedom by noticing that in the regions where the nonlinearities are effective all states are suppressed but the metastable ones (resonances). In this way the problem can be reformulated in terms of ordinary differential equations describing the evolution in the nonlinear regions weakly coupled to standard Schrödinger equations describing the propagation in the linear regions. We shall illustrate this idea, which we think is novel, in the prototype case of resonant tunneling through a double barrier. Our approach, beside providing an important mathematical advance, allows a better understanding of the physics involved. In particular we explain why the oscillations discovered in [2] disappear in the limit of very small and very large spacial width of the incoming cloud of electrons while they are maximal in intermediate situations. From the mathematical standpoint our discussion will be informal and computer assisted. However we think that a substantial part of it can be made completely rigorous. It is interesting to remark that the computer time needed for the numerical solution of the problem is reduced by several order of magnitudes with excellent agreement with the exact results.

Now we briefly recall the physical origin of our model problem. Interaction among the electrons can play a crucial role in electrical transport properties of mesoscopic systems [4]. As an example, let us consider a cloud of electrons moving in a double barrier heterostructure in which the well region confined between two potential barriers acts like a capacitor whose energy changes according to the electron charge trapped in it. We emphasize that the localization of the interaction is justified due to the existence of a resonance state which permits a long sojourn time inside the double barrier and therefore an accumulation of charge.

The exact time evolution of the depicted system can be written in terms of the anticommuting field operator associated to the electron cloud

$$i\hbar \frac{\partial}{\partial t}\hat{\Psi}(\vec{r},t) = \left[-\frac{\hbar^2}{2m}\nabla^2 + V(x) + \int \frac{e^2}{\varepsilon|\vec{r}-\vec{r}\,'|}\hat{\Psi}^\dagger(\vec{r}\,',t)\hat{\Psi}(\vec{r}\,',t)\,d\vec{r}\,'\right]\hat{\Psi}(\vec{r},t) \qquad (1)$$

where the external potential $V(x) = V_0\left(1_{[a,b]} + 1_{[c,d]}\right)$ with $a < b < c < d$ depends only on the coordinate $x$ orthogonal to the interfaces. As argued in [2] we can assume a decoupling between the degrees of freedom parallel and orthogonal to the interfaces and set up an approximate description in terms of a Hartree equation for a one particle wave function which depends only on the orthogonal coordinate. More precisely we factorize the single particle mean field in the following way

$$\Psi(\vec{r},t) \equiv \langle \hat{\Psi}(\vec{r},t) \rangle \simeq u(x,t)\phi(y,z)e^{-\frac{i}{\hbar}E_\parallel t} \quad . \tag{2}$$

where $\langle \ldots \rangle$ means expectation in the many body state at time $t$; $\phi(y,z)$ is a solution of the free particle Schrödinger equation in the plane parallel to the interfaces normalized to unity. We consider the interaction term effective only inside the well $[b,c]$ and we approximate it by the expression

$$\int \frac{e^2}{\varepsilon|\vec{r}-\vec{r}\,'|}|u(x',t)|^2|\phi(y',z')|^2 \, d\vec{r}\,' \simeq \alpha Q[u] 1_{[b,c]} \quad . \tag{3}$$

The constant $\alpha$ is $e^2/C$ where $C$ summarizes the result of the integration on the plane $yz$ and an averaging of the potential over the width of the well. Dimensionally $C$ is a length and defines the capacitance of the double barrier;

$$Q[u] = \int_b^c |u(x,t)|^2 \, dx \tag{4}$$

is the adimensional charge trapped in the well $[b,c]$ at time $t$. Within the above approximations we get the following one-dimensional nonlinear Schrödinger equation

$$i\hbar\frac{\partial}{\partial t}u(x,t) = \left[-\frac{\hbar^2}{2m}\frac{\partial^2}{\partial x^2} + V(x) + \alpha Q[u] 1_{[b,c]}\right] u(x,t) \tag{5}$$

which describes the dynamics of our quantum capacitor. An initial condition for $u(x,0)$ has to be given. We suppose $u(x,0)$ to be a Gaussian wave packet moving toward the double barrier and initially centered far from the well $[b,c]$ so that $Q \simeq 0$ at $t=0$.

Equation (5) has two conserved quantities, namely the number of particles

$$N = \int_{-\infty}^{+\infty} \overline{u(x,t)} u(x,t) \, dx \tag{6}$$

and the energy

$$E = \int_{-\infty}^{+\infty} \overline{u(x,t)} \left[-\frac{\hbar^2}{2m}\frac{\partial^2}{\partial x^2} + V(x)\right] u(x,t) \, dx + \frac{1}{2}\alpha Q[u]^2 \quad . \tag{7}$$

Some further comments are in order. The possibility of nonlinear effects in resonant tunneling was first envisaged in a paper by Ricco and Azbel [5] on the basis of a simple reaction-dissipation argument. The picture which emerges from [2] and the present study, as well as from a recent study of Malomed and Azbel [6] based on a different approach, is considerably more complex.

The time scale of these effects is in principle within the reach of present experimental techniques. An experimental verification would be of special interest because as shown in [7] this type of concentrated nonlinearities can induce a genuine chaotic behavior in heterostructures which does not have an obvious classical counterpart.

Equation (5) belongs to a class of equations which have other remarkable mathematical properties [8] beside those discussed in the present paper.

## II. THE ONE-MODE APPROXIMATION

In the following we will use the atomic unit system with $\hbar = 2m = 1$. In that case lengths are given in Bohr radii, 1 $a_0 = 0.529 \cdot 10^{-10}$ m, and energies in Rydbergs, 1 Ry= 13.6 eV. The unit of time is $4.83 \cdot 10^{-17}$ s.

We are interested in the initial value problem

$$\left(D_t + D_x^2 + V(x) + \alpha Q[u]\, 1_{[b,c]}\right) u(x,t) = 0 \tag{8a}$$

$$Q[u] = \|u\|_{[b,c]}^2 = \int_b^c |u(x,t)|^2\, dx \tag{8b}$$

$$u(x,0) = \beta e^{-\frac{1}{2}(x-x_0)^2 \sigma^{-2} + ik_0 x} \tag{8c}$$

where $D_t \equiv -i\frac{\partial}{\partial t}$, $D_x \equiv -i\frac{\partial}{\partial x}$ and with $V(x) = V_0 \left(1_{[a,b]} + 1_{[c,d]}\right)$ We consider $\alpha \geq 0$, $\beta > 0$, $\sigma > 0$, $k_0 > 0$, $a - x_0 \gg \sigma$ and $k_0^2$ close to the real part of a resonance $\lambda(0) = E_R(0) - i\Gamma(0)/2$ with $0 < E_R(0) < V_0$ and $\Gamma(0) > 0$, of the stationary Schrödinger operator $D_x^2 + V(x)$ on $\mathcal{R}$.

In this section we shall describe a one-mode approximation to problem (8) which hopefully can be rigorously justified in the limit $\sigma$, $d - c \to \infty$ when $V_0$ and $c - b$ are fixed and $\alpha$, $\beta$ are not too large, and provided that $\lambda(0)$ is a resonance so that $E_R(0)$ tends to some eigenvalue smaller than $V_0$ of the potential $V_0 \left(1_{]-\infty,b]} + 1_{[c,\infty[}\right)$ and $\Gamma(0) \to 0$ when $d - c \to \infty$. For simplicity we shall also assume that $E_R(0)$ is the only eigevalue of this potential at energy smaller than $V_0$.

We observe that the problem (8) is formally a Schrödinger equation with a time dependent potential. The standard way to study this situation is to work with the instantaneous eigenstates and eigenvalues for the associated time dependent Hamiltonian $H = D_x^2 + V(x) + \alpha Q 1_{[b,c]}$ [9]. One says that the $k$-th instantaneous eigenstate evolves adiabatically when transitions from it to any other $m$-th eigenstate cannot occur. This happens when [9]

$$\frac{\left|\dot{H}_{km}\right|}{(E_k - E_m)^2} \ll 1 \tag{9}$$

where $\dot{H}_{km}$ is the matrix element of the operator $\partial_t H$ between the $k$-th and the $m$-th eigenstates which have eigenvalues $E_k$ and $E_m$. In the problem (8) the instantaneous spectrum consists of a resonant state below the potential barrier $V_0$ and a continuum band above $V_0$. We are interested to make a one-mode adiabatic approximation by keepeng only the instantaneous resonant state in the description of the wave function inside the double barrier. Since the distance between the resonant state and the rest of the spectrum is of the order of $V_0$, we get the adiabatic condition

$$\frac{\alpha \left|\dot{Q}\right|}{V_0^2} \ll 1\ . \tag{10}$$

We shall verify a posteriori that this condition is satisfied in the range of parameters we use.

We now derive the equation for the one-mode approximation. Let $\tilde{V}(x) = V_0 \, 1_{[a,d]}$ be the potential obtained from $V(x)$ by filling the potential well $[b,c]$. We start by solving the initial value problem

$$\left(D_t + D_x^2 + \tilde{V}(x)\right) \tilde{\mu}(x,t) = 0 \tag{11a}$$

$$\tilde{\mu}(x,0) = u(x,0) \tag{11b}$$

and look for a solution of Eq. (8) of the form $u = \tilde{\mu} + v$. Then Eq. (8) becomes

$$\left(D_t + D_x^2 + V(x) + \alpha Q[\tilde{\mu} + v] \, 1_{[b,c]}\right) v(x,t) = \left(\tilde{V}(x) - V(x) - \alpha Q[\tilde{\mu} + v] \, 1_{[b,c]}\right) \tilde{\mu}(x,t) \tag{12a}$$

$$Q[\tilde{\mu} + v] = \int_b^c |\tilde{\mu}(x,t) + v(x,t)|^2 \, dx \tag{12b}$$

$$v(x,0) = 0 \, . \tag{12c}$$

If $0 \leq s \leq V_0$, we let $\lambda(s) = E_R(s) - i\Gamma(s)/2$ be the resonance of $D_x^2 + V_s(x)$ where $V_s(x) = V(x) + s \, 1_{[b,c]}$. Let $e(s,x)$ be the corresponding resonance state

$$\left(D_x^2 + V_s(x) - \lambda(s)\right) e(s,x) = 0 \, . \tag{13}$$

We recall that $e(s,x)$ is a function on $\mathcal{R}$ such that $e(s,x) \propto \exp\left(\pm i\sqrt{\lambda(s)}x\right)$, for $x \to \pm\infty$. We also observe that, as a special case of the method of complex scaling [10], $e(s,x)$ is of class $L^2$ on the contour $\gamma = e^{-i\theta} \, ]-\infty, a] \cup [a,d] \cup e^{i\theta} \, [d, +\infty[$ if $\theta > 0$ is conveniently chosen. We normalize the resonance state to unity

$$\int_\gamma e(s,x)^2 \, dx = 1 \, . \tag{14}$$

In the spirit of the one-mode approximation, we assume $v(x,t) = \tilde{z}(t)e(s,x)$ with

$$s = \alpha Q[\tilde{\mu} + \tilde{z} e(s)] \, . \tag{15}$$

The problem (12) then becomes

$$\left(D_t + D_x^2 + V(x) + \alpha Q[\tilde{\mu} + \tilde{z}e] \, 1_{[b,c]}\right) \tilde{z}(t) e(s,x) = (V_0 - \alpha Q[\tilde{\mu} + \tilde{z}e]) 1_{[b,c]} \, \tilde{\mu}(x,t) \tag{16}$$

with initial condition $\tilde{z}(0) = 0$. If Eq. (15) defines a unique $s = s[\tilde{\mu}, \tilde{z}]$, Eq. (16) reduces to

$$(D_t + \lambda(s)) \tilde{z}(t) e(s,x) = (V_0 - s) \, 1_{[b,c]} \, \tilde{\mu}(x,t) \, . \tag{17}$$

We multiply the above equation by $e(s,x)$ and integrate with respect to $x$ on the contour $\gamma$. We also use the normalization condition for $e$ implying $\int_\gamma D_t\left[e(s,x)\right] e(s,x) \, dx = 0$ and we obtain

$$(D_t + \lambda(s))\,\tilde{z}(t) = (V_0 - s)\int_b^c \tilde{\mu}(x,t)e(s,x)\,dx\ . \tag{18}$$

Introducing $\mu(x,t) = e^{iE_0 t}\tilde{\mu}(x,t)$ and $z(t) = e^{iE_0 t}\tilde{z}(t)$, where $E_0 = k_0^2$, we can rewrite Eq. (15) and Eq. (18) as

$$\frac{dz(t)}{dt} = -i\left(\lambda(s) - E_0\right)z(t) + i(V_0 - s)\int_b^c \mu(x,t)e(s,x)\,dx \tag{19a}$$

$$s = \alpha Q[\mu + ze] \tag{19b}$$

$$z(0) = 0\ . \tag{19c}$$

The above three equations represent our one-mode approximation. Since $s$ depends on $z(t)$ when $\alpha \neq 0$, Eq. (19) is clearly a nonlinear ordinary differential equation with a driving term.

### A. The driving term

In principle a complete integral representation of the driving term $\tilde{\mu}(x,t)$ or $\mu(x,t)$ could be given, but we prefer to use various approximations which seem justified in the limit $\sigma$, $d - a \to \infty$.

We start by looking for a solution of the stationary problem $\left(D_x^2 + \tilde{V}(x) - E\right)\tilde{e}(k,x) = 0$ of the form

$$\tilde{e}(k,x) = \frac{1}{\sqrt{2\pi}}\begin{cases} e^{ikx} + r(k)e^{-ikx} & x < a \\ g^+(k)e^{\kappa x} + g^-(k)e^{-\kappa x} & a < x < d \\ t(k)e^{ikx} & d < x \end{cases} \tag{20}$$

where $k = \sqrt{E}$ and $\kappa = \sqrt{V_0 - k^2}$. The normalization constant is chosen in such a way that $\int \overline{\tilde{e}(k,x)}\tilde{e}(k',x)\,dx = \delta(k-k')$. The functions $r(k)$, $t(k)$ and $g^\pm(k)$ can be evaluated exactly by requiring $\tilde{e}$ to be of class $C^1$. In the limit of $\kappa(d-a) \gg 1$ we get the simple semiclassical result

$$r(k) = \frac{ik+\kappa}{ik-\kappa}e^{2ika} + \mathcal{O}\left(e^{-\kappa(d-a)}\right) \tag{21}$$

$$g^-(k) = \frac{2ik}{ik-\kappa}e^{(ik+\kappa)a} + \mathcal{O}\left(e^{-\kappa(d-a)}\right) \tag{22}$$

$$t(k),\ g^+(k) = \mathcal{O}\left(e^{-\kappa(d-a)}\right)\ . \tag{23}$$

Now we have the solution of Eq. (11) of the form

$$\tilde{\mu}(x,t) = \int_0^\infty f(k)e^{-ik^2 t}\,\tilde{e}(k,x)\,dk \tag{24}$$

where, in the limit $k_0\sigma \gg 1$, $f(k)$ is the Fourier transform of $\tilde{\mu}(x,0)$

$$f(k) = \beta\sigma e^{-\frac{1}{2}(k_0-k)^2\sigma^2 + i(k_0-k)x_0} . \tag{25}$$

The integral in Eq. (24) can be evaluated analytically by extending the lower limit to $-\infty$ and approximating the smooth functions of $k$ with their value at $k = k_0$. The integration of the remaining quadratic exponential gives for $a < x < d$

$$\tilde{\mu}(x,t) = \frac{2ik_0 e^{-\kappa_0(x-a)}}{ik_0 - \kappa_0} \frac{\beta}{\sqrt{1+2it\sigma^{-2}}} \exp\left\{\frac{[k_0\sigma + i(a-x_0)\sigma^{-1}]^2}{2(1+2it\sigma^{-2})} - \frac{1}{2}k_0^2\sigma^2 + ik_0x_0\right\} \tag{26}$$

where $\kappa_0 = \sqrt{V_0 - k_0^2}$. This expression further simplifies for small times or large $\sigma$. By expanding with respect to $t\sigma^{-2}$ and using $k_0\sigma \gg 1$ we get to the leading order

$$\mu(x,t) = \frac{2i\beta k_0 e^{-\kappa_0(x-a)+ik_0 a}}{ik_0 - \kappa_0} e^{-\frac{1}{2}\left(\frac{\Gamma_0}{2}\right)^2 (t-t_0)^2} \tag{27}$$

where $t_0 = (a - x_0)/(2k_0)$ is the time a classical particle with kinetic energy $E_0$ spends to cover the distance $a - x_0$ and $\Gamma_0 = 2\sqrt{2}k_0/\sigma$ is the full width at height $e^{-1/2}$ of $|f(k)|^2$, i.e. the energy spread of the initial wave function $u(x,0)$.

### B. Solution of the one-mode approximation in the linear case

We limit ourselves to a qualitative discussion of the limiting behaviors for $\Gamma_0 \ll \Gamma(0)$ and $\Gamma_0 \gg \Gamma(0)$. Since $\alpha = 0$, Eq. (19) becomes

$$\frac{dz(t)}{dt} = -i(\lambda(0) - E_0)z(t) + iV_0 \int_b^c \mu(x,t)e(0,x)\,dx \tag{28}$$

with initial condition $z(0) = 0$. In the resonant case $E_0 = E_R(0)$ we get

$$\frac{dz(t)}{dt} + \frac{\Gamma(0)}{2}z(t) = iV_0 \int_b^c \mu(x,t)e(0,x)\,dx . \tag{29}$$

The driving term in the right hand side has been evaluated in the previous section. It has a roughly costant argument and a modulus which behaves like $\exp\left(-\frac{1}{2}(\Gamma_0/2)^2(t-t_0)^2\right)$ at least for $\sigma$ large. In this case if $\Gamma_0 \ll \Gamma(0)$ the Gaussian right hand side of Eq. (29) varies very little during the characteristic time period $2/\Gamma(0)$ of the left hand side and we get the approximate solution

$$z(t) \simeq \frac{2iV_0}{\Gamma(0)} \int_b^c \mu(x,t)e(0,x)\,dx \tag{30}$$

which has the same shape as the driving term, i.e. a Gaussian shape. On the other hand if $\Gamma_0 \gg \Gamma(0)$ we get the approximate solution

$$z(t) \simeq iV_0 e^{-\frac{\Gamma(0)}{2}t} \int_0^t \int_b^c \mu(x,t')e(0,x)\,dx\,dt' \tag{31}$$

characterized by a linear exponential decay slow with respect to the rise time.

The above results can be generalized to the non resonant case. We get

$$z(t) \simeq \frac{V_0}{E_R(0) - E(0) - i\Gamma(0)/2} \int_b^c \mu(x,t) e(0,x) \, dx \tag{32}$$

when $\Gamma_0 \ll \Gamma(0)$ and

$$z(t) \simeq iV_0 e^{-i\left(E_R(0) - E(0) - i\frac{\Gamma(0)}{2}\right)t} \int_0^t \int_b^c \mu(x,t') e(0,x) \, dx \, dt' \tag{33}$$

when $\Gamma_0 \gg \Gamma(0)$.

It is worth noting the conceptual simplicity of the one-mode approach in this case if compared with previous treatments [11].

### C. Numerical simulations

We now introduce the numerical simulations of the one-mode approximation obtained by a standard Runge-Kutta integration of Eq. (19). For comparison we give also the numerical simulations of the Eq. (8) obtained with the algorithm described in Appendix B.

We consider an external potential $V(x)$ with $c - b = 15$, $d - c = b - a = 20$ and $V_0 = 0.3/13.6$. In this case we have a resonace $\lambda(0)$ with $E_R(0) \simeq 0.5 \, V_0$ and $\Gamma_0 \simeq 5 \cdot 10^{-3}/13.6$. The initial wave function is chosen to be in resonance with $k_0^2 = E_R(0)$. We consider three different values for its width: $\sigma = 5775$, $\sigma = 825$ and $\sigma = 110$ corresponding to $\Gamma_0 \simeq 0.14 \, \Gamma(0)$, $\Gamma_0 \simeq \Gamma(0)$ and $\Gamma_0 \simeq 7.3 \, \Gamma(0)$, respectively. A constant $\beta = (825\sqrt{\pi})^{-1/2}$ is used is every case. Finally we put $a - x_0 = 5\sigma$ so that no appreciable charge $Q$ is present in the well at $t = 0$. Notice that the semiclassical condition and the condition $k_0\sigma \gg 1$, employed in evaluating the approximate expression (26) for the driving term, are well satisfied with the above parameters.

Fig. 1, 2 and 3 show the behavior of the charge $Q(t)$ for the three given values of $\sigma$ and for different values of the parameter $\alpha$. Fig. 4, 5 and 6 show the trajectories of $z(t)$ in the complex plane for the same cases. We notice that the one-mode approximation reproduces exceptionally well the exact result. A small difference appears and grows up when $\sigma$ is decreased and $\alpha$ is increased in agreement with the adiabatic condition (10). This point will be discussed in detail in Section IV.

The limiting behaviors for $\Gamma_0 \ll \Gamma(0)$ and $\Gamma_0 \gg \Gamma(0)$ in the linear case ($\alpha = 0$) as expressed by Eq. (30) and Eq. (31) are apparent in Fig. 1 and Fig. 3. To understand qualitatively the nonlinear behavior we introduce a simplified one-mode approximation which keeps the basic features of Eq. (19) and is analytically more manageable.

### III. THE SIMPLIFIED ONE-MODE APPROXIMATION

If we assume that Eq. (19b) has a unique solution $s = s[\mu + ze]$, then the one-mode approximation (o.m.a.) defined by Eq. (19) can be written

$$\dot{z} = \mathcal{V}(t, z(t)) \tag{34}$$

where

$$\mathcal{V}(t,z) = -i\left(\lambda(s[\mu+ze]) - E_0\right)z + i(V_0 - s[\mu+ze])\int_b^c \mu(x,t)e(s[\mu+ze],x)\,dx\ . \quad (35)$$

It turns out that the qualitative numerical aspects of the vector field $\mathcal{V}$ are already captured by a simplified one-mode approximation (s.o.m.a.). The s.o.m.a. is obtained from the o.m.a. *i)* replacing $e(s,x)$ everywhere by $e(0,x)$, *ii)* replacing $V_0 - s$ by $V_0$, *iii)* replacing $Q[\mu+ze]$ by $Q[ze]$ and *iv)* replacing $\lambda(s) - E_0$ by $E_R'(0)s - i\Gamma(0)/2$, where the derivative

$$E_R'(0) = \left.\frac{dE_R(s)}{ds}\right|_{s=0} = 1 - \frac{E_R(0)\cos^2\left[\sqrt{E_R(0)}(c-b)/2\right]}{V_0\cos^2\left[\sqrt{E_R(0)}(c-b)/2\right] + E_R(0)\sqrt{V_0 - E_R(0)}(c-b)/2} \quad (36)$$

is evaluated from Eq. (A4). The condition in Eq. (19b) now simplifies to

$$s = \alpha q|z|^2 \qquad q = \int_b^c |e(0,x)|^2 dx \quad (37)$$

and the vector field becomes

$$\mathcal{V}(t,z) = -\left(\frac{\Gamma(0)}{2} + iE_R'(0)\alpha q|z|^2\right)z + iV_0\int_b^c \mu(x,t)e(0,x)\,dx\ . \quad (38)$$

According to Eq. (27) during the interesting part of the evolution the driving term is well approximated by

$$iV_0\int_b^c \mu(x,t)e(0,x)\,dx = F\ e^{-\frac{1}{2}\left(\frac{\Gamma_0}{2}\right)^2(t-t_0)^2} \quad (39)$$

where

$$F = \frac{2\beta k_0 V_0 e^{ik_0 a}}{\kappa_0 - ik_0}\int_b^c e^{-\kappa_0(x-a)}e(0,x)\,dx\ . \quad (40)$$

### A. solution of the simplified one-mode approximation

We consider first two limiting behaviors in analogy to the discussion of the linear case in Section II B.

When $\Gamma_0 \ll \Gamma(0) + E_R'(0)\alpha q|z|^2$ it is $\dot{z}(t) \simeq 0$ and $z(t)$ is close to the instantaneous fixed point $z_c(t)$ defined by $\mathcal{V}(t,z_c(t)) = 0$. If $z_c(t)$ is such a fixed point we have

$$\left(\frac{\Gamma(0)^2}{4} + E_R'(0)^2\alpha^2 q^2|z_c|^4\right)|z_c|^2 = V_0^2\left|\int_b^c \mu(x,t)e(0,x)\,dx\right|^2 \quad (41)$$

and conversely if $|z_c|$ solves Eq. (41) then we get the unique fixed point of modulus $|z_c|$

$$z_c = \frac{iV_0\int_b^c \mu(x,t)e(0,x)\,dx}{\Gamma(0)/2 + iE_R'(0)\alpha q|z_c|^2}\ . \quad (42)$$

Since the left hand side of Eq. (41) is a strictly increasing function of $|z_c|$ it is clear that $\mathcal{V}$ has a unique fixed point given by Eq. (41) and Eq. (42). In this situation we do not have oscillations in $|z_c(t)|$ and therefore in $Q(t)$ as shown in Fig. 1.

Independently on the validity of the previous inequality, we distinguish two zones for the fixed point behavior. *a) the linear zone* : $E'_R(0)\alpha q|z_c|^2 \ll \Gamma(0)/2$. In this zone we have

$$z_c(t) \simeq \frac{2iV_0 \int_b^c \mu(x,t)e(0,x)\,dx}{\Gamma(0)} \ . \tag{43}$$

Using this expression the linear zone is defined by the condition

$$8E'_R(0)\alpha q|F|^2 e^{-\left(\frac{\Gamma_0}{2}\right)^2(t-t_0)^2} \ll \Gamma(0)^3 \ . \tag{44}$$

Due to the time dependence of the driving term we are sure to be in the linear zone at the beginning and at the end of the process. *b) the nonlinear zone*: $E'_R(0)\alpha q|z_c|^2 \gg \Gamma(0)$. In this zone from Eq. (41) we have

$$|z_c(t)| \simeq \left|\frac{V_0 \int_b^c \mu(x,t)e(0,x)\,dx}{E'_R(0)\alpha q}\right|^{1/3} \tag{45}$$

and from Eq. (42)

$$\arg z_c(t) \simeq \arg \int_b^c \mu(x,t)e(0,x)\,dx \ . \tag{46}$$

Therefore the nonlinear zone is defined by

$$(E'_R(0)\alpha q)^{1/3}|F|^{2/3} e^{-\frac{1}{3}\left(\frac{\Gamma_0}{2}\right)^2(t-t_0)^2} \gg \Gamma(0) \ . \tag{47}$$

Notice that the argument of $z_c$ changes by $\pi/2$ when we go from the linear to the nonlinear zone. When the driving term reaches its maximum, $z_c(t)$ becomes approximatively stationary and then returns to the origin following a curve close to the outgoing trajectory. This kind of behavior is well illustrated in Fig. 4 and 5.

When $\Gamma_0 \gg \Gamma(0) + E'_R(0)\alpha q|z|^2$ we have the other limiting behavior characterized by the approximate equation

$$z(t) \simeq iV_0 e^{-\frac{\Gamma(0)}{2}t - iE'_R(0)\alpha q \int_0^t |z(t')|^2\,dt'} \int_0^t \int_b^c \mu(x,t')e(0,x)\,dx\,dt' \tag{48}$$

and therefore

$$|z(t)|^2 = |F|^2 e^{-\Gamma(0)t}\left(\int_0^t e^{-\frac{1}{2}\left(\frac{\Gamma_0}{2}\right)^2(t'-t_0)^2}\,dt'\right)^2 \tag{49}$$

which is as in the linear case without oscillations in $Q(t)$. The corresponding condition can be written more explicitely as

$$\Gamma_0 \gg \Gamma(0) + E'_R(0)\alpha q|F|^2 e^{-\Gamma(0)t}\left(\int_0^t e^{-\frac{1}{2}\left(\frac{\Gamma_0}{2}\right)^2(t'-t_0)^2}\,dt'\right)^2 \ . \tag{50}$$

The above condition is well satisfied in the simulation of Fig. 3 corresponding to $\alpha = 0.05$ while it is violated for $\alpha = 5$. In this case oscillations of $Q(t)$ appear. In the case $\alpha = 0.5$ the two sides of Eq. (50) are of the same order. From Fig. 6 we see that the trajectory of $z(t)$ is always far from the trajectory of $z_c(t)$. Still $q|z_c(t)|^2$ gives a good estimate of the average, i.e. non oscillating, value of $Q(t)$.

For intermediate situations, that is $\Gamma_0 \simeq \Gamma(0)$, the analytical treatment is necessarily more qualitative. In this connection it is convenient to introduce the linearized solution $z_l(t)$ around the fixed point $z_c(t)$

$$\frac{d}{dt}\begin{pmatrix} z_l \\ \overline{z_l} \end{pmatrix} = \mathcal{L} \begin{pmatrix} z_l - z_c \\ \overline{z_l} - \overline{z_c} \end{pmatrix} \tag{51}$$

where

$$\mathcal{L} = \begin{pmatrix} \frac{\partial \mathcal{V}}{\partial z}(z_c) & \frac{\partial \mathcal{V}}{\partial \overline{z}}(z_c) \\ \frac{\partial \overline{\mathcal{V}}}{\partial \overline{z}}(z_c) & \frac{\partial \overline{\mathcal{V}}}{\partial z}(z_c) \end{pmatrix} . \tag{52}$$

The linearization matrix $\mathcal{L}$ has instantaneous eigenvalues

$$\ell_\pm = \Re e \frac{\partial \mathcal{V}}{\partial z}(z_c) \pm i\sqrt{\left(\Im m \frac{\partial \mathcal{V}}{\partial z}(z_c)\right)^2 - \left|\frac{\partial \mathcal{V}}{\partial \overline{z}}(z_c)\right|^2} \tag{53}$$

which in the s.o.m.a. are given by

$$\ell_\pm = -\frac{\Gamma(0)}{2} \pm i\sqrt{3} E'_R(0)\alpha q|z_c|^2 . \tag{54}$$

In the linear zone the real part dominates in this expression while in the nonlinear zone it is the imaginary part that dominates. In every case the fixed point is attractive since $\Re e\, \ell_\pm < 0$. In the nonlinear zone we have an important rotation around the fixed point with angular speed

$$\Im m\, \ell_\pm = \sqrt{3} \left(E'_R(0)\alpha q\right)^{1/3} |F|^{2/3} \tag{55}$$

while the purely attractive behavior dominates in the linear zone. It is also possible to see that the rotation is oriented clock-wise.

In Fig. 7 we give the result of the simulation of the s.o.m.a. for $\sigma = 825$ and $\alpha = 5$ and we compare it with the linearization around the corresponding fixed point. We see a qualitative agreement with the o.m.a. result in Fig. 5. The linearization around the fixed point provides a good insight into the structure of the oscillating behavior.

## IV. ADIABATIC CONDITION

In this section we give an explicit expression of the adiabatic condition (10) in terms of the relevant parameters. We use the s.o.m.a. with $Q(t) = q|z(t)|^2$ and we situate our discussion in the most interesting region where $|t - t_0|\Gamma_0/2 \leq 1$. We approximate $d|z(t)|^2/dt \simeq d|z_c(t)|^2/dt$. This is not entirely correct because, due to the oscillations, the derivative of $|z(t)|^2$ may reach higher values. However, and this is confirmed by numerical simulations, the order of magnitude of the maximum value of $d|z(t)|^2/dt$ is given by the slope of the first oscillation which is close to the maximum value of $d|z_c(t)|^2/dt$. In the interesting nonlinear zone the value of $|z_c(t)|^2$ is given by Eq. (45) and its time derivative is evaluated by using Eq. (39) for the driving term

$$\left|\frac{d|z_c(t)|^2}{dt}\right| \lesssim \Gamma_0 \frac{|F|^{2/3}}{\alpha^{2/3}} \tag{56}$$

where we used $E'_R(0)q \simeq 1$. Noticing that $e(0,x) \sim (c-b)^{-1/2}$ in $[b,c]$, from Eq. (40) we get

$$|F| \sim \frac{2\beta k_0 \sqrt{V_0} e^{-\kappa_0(b-a)}}{\kappa_0 \sqrt{c-b}} . \tag{57}$$

The adiabatic condition (10) becomes

$$\alpha^{1/3} \frac{\Gamma_0}{V_0^2} \left(\frac{2\beta k_0 \sqrt{V_0} e^{-\kappa_0(b-a)}}{\kappa_0 \sqrt{c-b}}\right)^{2/3} \ll 1 . \tag{58}$$

We make this inequality physically more transparent by inserting the width of the resonance. From Eq. (A9) we have

$$\Gamma(0) \sim \frac{\kappa_0^2 e^{-2\kappa_0(b-a)}}{k_0(c-b)} \tag{59}$$

and, since in our case $k_0 \sim \kappa_0$, the adiabatic condition (10) is also

$$\frac{\Gamma_0}{V_0}\left(\frac{\alpha}{V_0}\frac{\Gamma(0)}{V_0}\sqrt{\frac{\beta^4}{V_0}}\right)^{1/3} \ll 1 . \tag{60}$$

The order of magnitude of the left hand side is in good agreement with the maximal value of $\alpha \left|\dot{Q}\right| V_0^2$ as obtained from numerical simulations and the inequality is well satisfied. In fact the left hand side ranges from $10^{-3}$ in the case $\sigma = 5775$ to $3 \cdot 10^{-2}$ in the case $\sigma = 110$ for $\alpha = 5$.

### ACKNOWLEDGMENTS

This collaboration has been made possible by the support of CNRS France, INFN Italy, Mittag-Leffler Institute Sweden and the Commission of European Communities under contract SC1-CT91-0695. It is a pleasure to thank Federico Capasso for stimulating discussions during the development of this work.

### APPENDIX A: EIGENVALUE AND EIGENFUNCTION AT RESONANCE

We want to calculate the eigenvalue $\lambda(s) = E_R(s) - i\Gamma(s)/2$ and the eigenfunction $e(s,x)$ of the stationary problem

$$\left(D_x^2 + V_s(x) - \lambda(s)\right) e(s,x) = 0 \tag{A1}$$

with $V_s(x) = V_0\, 1_{[a,b]} + s\, 1_{[b,c]} + V_0\, 1_{[c,d]}$ and $0 < s < V_0$. The eigenfunction $e(s,x)$ has to be known only in the interval $[b,c]$. For convenience we choose $a = -d$ and $b = -c$.

Working with the assumption that $d - c$ is large, we approximate $E_R(s)$ by the first eigenvalue $E$ for the confining potential $V_{s,\infty}(x) = V_0\, 1_{]-\infty,b]} + s\, 1_{[b,c]} + V_0\, 1_{[c,\infty[}$ and we approximate the resonant state $e(s,x)$ in $[b,c]$ by the corresponding eigenfunction. Observing that $e(s,x)$ is an even function of $x$, up to a normalization constant we have

$$e(s,x) = \begin{cases} \cos(\sqrt{E-s}\, x) & 0 < x < c \\ \gamma e^{-\sqrt{V_0-E}\,(x-c)} & c < x \end{cases}. \tag{A2}$$

The $C^1$ requirement at $x = c$ gives

$$\gamma = \cos(\sqrt{E-s}\, c) \tag{A3}$$

and

$$\frac{\sqrt{V_0-E}}{\sqrt{E-s}} = \tan\left(\sqrt{E-s}\, c\right). \tag{A4}$$

This equation has a unique solution $s < E < V_0$ which can be found numerically. Once the solution $E$ is known, we normalize the corresponding eigenfunction and we get

$$e(s,x) = \frac{\cos(\sqrt{E-s}\, x)}{\sqrt{c + \frac{\sin(2\sqrt{E-s}\, c)}{2\sqrt{E-s}} + \frac{1+\cos(2\sqrt{E-s}\, c)}{2\sqrt{V_0-E}}}} \tag{A5}$$

in the interesting interval $0 < x < c$.

Finally we derive a semiclassical formula for $\Im m\, \lambda(s)$. In this case we improve the above approximation for $e(s,x)$ valid only for $|x| < c$ by considering the reflected waves in the potential $V_s(x)$. In the semiclassical limit up to a normalization constant we can write

$$e(s,x) = \begin{cases} \cos(\sqrt{E-s}\, x) & 0 < x < c \\ \gamma e^{-\sqrt{V_0-E}\,(x-c)} + \delta e^{-\sqrt{V_0-E}\,(d-c)+\sqrt{V_0-E}\,(x-d)} & c < x < d \\ \eta e^{i\sqrt{E}\,(x-d)} & d < x \end{cases}. \tag{A6}$$

The costant $\gamma$ is given by Eq. (A3) while $\delta$ and $\eta$ can be obtained by $C^1$ requirement at $x = d$. In particular we have

$$\eta = \frac{2\sqrt{V_0-E}}{\sqrt{V_0-E} - i\sqrt{E}}\cos(\sqrt{E-s}\, c)\, e^{-\sqrt{V_0-E}\,(d-c)}. \tag{A7}$$

By using Eq. (A1) and its complex conjugate we can write

$$\left[\lambda(s) - \overline{\lambda(s)}\right] \int_{-\widetilde{d}}^{\widetilde{d}} e(s,x)\, \overline{e(s,x)}\, dx =$$

$$\int_{-\widetilde{d}}^{\widetilde{d}} e(s,x)\, \partial_x^2 \overline{e(s,x)}\, dx - \int_{-\widetilde{d}}^{\widetilde{d}} \partial_x^2 e(s,x)\, \overline{e(s,x)}\, dx = 4i\Im m\left[e(s,\widetilde{d})\, \partial_x\overline{e(s,\widetilde{d})}\right] \tag{A8}$$

where $\tilde{d} - d > 0$ is small. Inserting the expression (A6) for $e(s,x)$ at $x = \tilde{d}$ we get

$$\Gamma(s) = \frac{16\sqrt{E}(V_0 - E)\cos^2(\sqrt{E-s}\ c)\ e^{-2\sqrt{V_0-E}\ (d-c)}}{V_0 \left[c + \frac{\sin(2\sqrt{E-s}\ c)}{2\sqrt{E-s}} + \frac{1+\cos(2\sqrt{E-s}\ c)}{2\sqrt{V_0-E}}\right]} \tag{A9}$$

where $E = E_R(s)$ has been evaluated above.

## APPENDIX B: NUMERICAL ALGORITHM

Here we discuss briefly the algorithm used for the numerical simulations of Eq. (8). The partial differential equation is transformed in a set of algebraic equations by discretizing the space-time $x - t$ in an appropriate two dimensional lattice. Let us suppose that we want to follow the temporal evolution of the initial state $u(x, 0)$ during the time interval $[0, t_{max}]$. Let $[x_{min}, x_{max}]$ be the space interval where the wavefunction $u(x, t)$ can be considered localized for $0 \leq t \leq t_{max}$. The space-time lattice is defined by putting

$$x \to x_{min} + j\Delta x \qquad j = 0, 1, 2, \ldots, J+1 \tag{B1}$$

$$t \to n\Delta t \qquad n = 0, 1, 2, \ldots, N \tag{B2}$$

where the total number of points in the lattice is related to the lattice constants $\Delta x$ and $\Delta t$ by

$$(J+1)\Delta x = x_{max} - x_{min} \tag{B3}$$

$$N\Delta t = t_{max} \ . \tag{B4}$$

The wave function reduces on the lattice to a matrix

$$u(x, t) \to u_j^n \tag{B5}$$

where $u_j^0$ are kown. The Hamiltonian operator $H = D_x^2 + W$, where $W = V + \alpha Q 1_{[b,c]}$ includes both the external and the nonlinear potentials, becomes

$$Hu(x,t) \to (Hu)_j^n = -\Delta x^{-2}(u_{j+1}^n - 2u_j^n + u_{j-1}^n) + W_j^n u_j^n \tag{B6}$$

with $W_j^n = W(x_j, t_n)$.

The discretization of the operator $D_t$ is obtained with the Cayley approximation for the exponential time evolution operator

$$e^{-i\Delta tH} \simeq \left[1 + \frac{i}{2}\Delta tH\right]^{-1}\left[1 - \frac{i}{2}\Delta tH\right] \tag{B7}$$

which preserves unitarity. The finite difference scheme for time evolution gives the implicit equation

$$\left[1 + \frac{i}{2}\Delta t H\right] u(t + \Delta t) = \left[1 - \frac{i}{2}\Delta t H\right] u(t) \tag{B8}$$

which corresponds in the lattice to the following linear system

$$\omega u_{j-1}^{n+1} + \xi_j^n u_j^{n+1} + \omega u_{j+1}^{n+1} = \chi_j^n \tag{B9}$$

where

$$\chi_j^n = \overline{\omega} u_{j-1}^n + \overline{\xi_j^n} u_j^n + \overline{\omega} u_{j+1}^n \tag{B10}$$

and

$$\xi_j^n = 1 + i\frac{\Delta t}{\Delta x^2} + \frac{i}{2}\Delta t W_j^n \tag{B11}$$

$$\omega = -\frac{i}{2}\frac{\Delta t}{\Delta x^2} . \tag{B12}$$

By using the condition $u_0^n = u_{J+1}^n = 0$ which expresses the fact that the wavefunction is localized in $[x_{min}, x_{max}]$ we can write the following tridiagonal system

$$\begin{pmatrix} \xi_1^n & \omega & 0 & \ldots & 0 & 0 & 0 \\ \omega & \xi_2^n & \omega & \ldots & 0 & 0 & 0 \\ 0 & \omega & \xi_3^n & \ldots & 0 & 0 & 0 \\ \vdots & \vdots & \vdots & \ddots & \vdots & \vdots & \vdots \\ 0 & 0 & 0 & \ldots & \xi_{J-2}^n & \omega & 0 \\ 0 & 0 & 0 & \ldots & \omega & \xi_{J-1}^n & \omega \\ 0 & 0 & 0 & \ldots & 0 & \omega & \xi_J^n \end{pmatrix} \begin{pmatrix} u_1^{n+1} \\ u_2^{n+1} \\ u_3^{n+1} \\ \vdots \\ u_{J-2}^{n+1} \\ u_{J-1}^{n+1} \\ u_J^{n+1} \end{pmatrix} = \begin{pmatrix} \chi_1^n \\ \chi_2^n \\ \chi_3^n \\ \vdots \\ \chi_{J-2}^n \\ \chi_{J-1}^n \\ \chi_J^n \end{pmatrix} . \tag{B13}$$

Starting from $u_j^0$ the solution of the above system at each time step gives the unknown $u_j^{n+1}$ in terms of the known $u_j^n$.

Due to its tridiagonal nature the linear system in Eq. (B13) is efficently solved by standard triangularization methods requiring a number of operation $\mathcal{O}(J)$ with a total computation time $\mathcal{O}(JN)$. The number of lattice points $N$ and $J$ is chosen in such a way that $u_j^n$ is a good approximation to the continuous function $u(x,t)$. With the parameters given in Section II, the difference between $u_j^n$ and $u(x,t)$ is estimated to be less than 0.1 % for $\Delta x \leq 1$ and $\Delta t \leq 4$.

FIGURES

FIG. 1. Comparison between the charge $Q(t)$ obtained by numerical solution of the partial differential equation (8) (dot-dashed line) and of the one-mode approximation (19) (full line) as a function of the nonlinearity strength $\alpha$. Notice that the two curves almost coincide. The values of the parameters are given in the text. Here $\sigma = 5775$ which corresponds to $\Gamma_0 \simeq 0.14\, \Gamma(0)$.

FIG. 2. As in Fig. 1 but for $\sigma = 825$ which corresponds to $\Gamma_0 \simeq \Gamma(0)$.

FIG. 3. As in Fig. 1 but for $\sigma = 110$ which corresponds to $\Gamma_0 \simeq 7.3\, \Gamma(0)$.

FIG. 4. Trajectory of the numerical solution $z(t)$ of the one-mode approximation (19) for the same cases of Fig. 1 (full line). The dashed line is the trajectory of the corresponding fixed point $z_c(t)$.

FIG. 5. Trajectory of the numerical solution $z(t)$ of the one-mode approximation (19) for the same cases of Fig. 2 (full line). The dashed line is the trajectory of the corresponding fixed point $z_c(t)$.

FIG. 6. Trajectory of the numerical solution $z(t)$ of the one-mode approximation (19) for the same cases of Fig. 3 (full line). The dashed line is the trajectory of the corresponding fixed point $z_c(t)$.

FIG. 7. Trajectory of the numerical solution $z(t)$ of the simplified one-mode approximation for the case $\sigma = 825$, $\alpha = 5$ (full line). The dashed line is the trajectory of the corresponding fixed point $z_c(t)$ while the dotdashed line is the linearized solution $z_l(t)$.

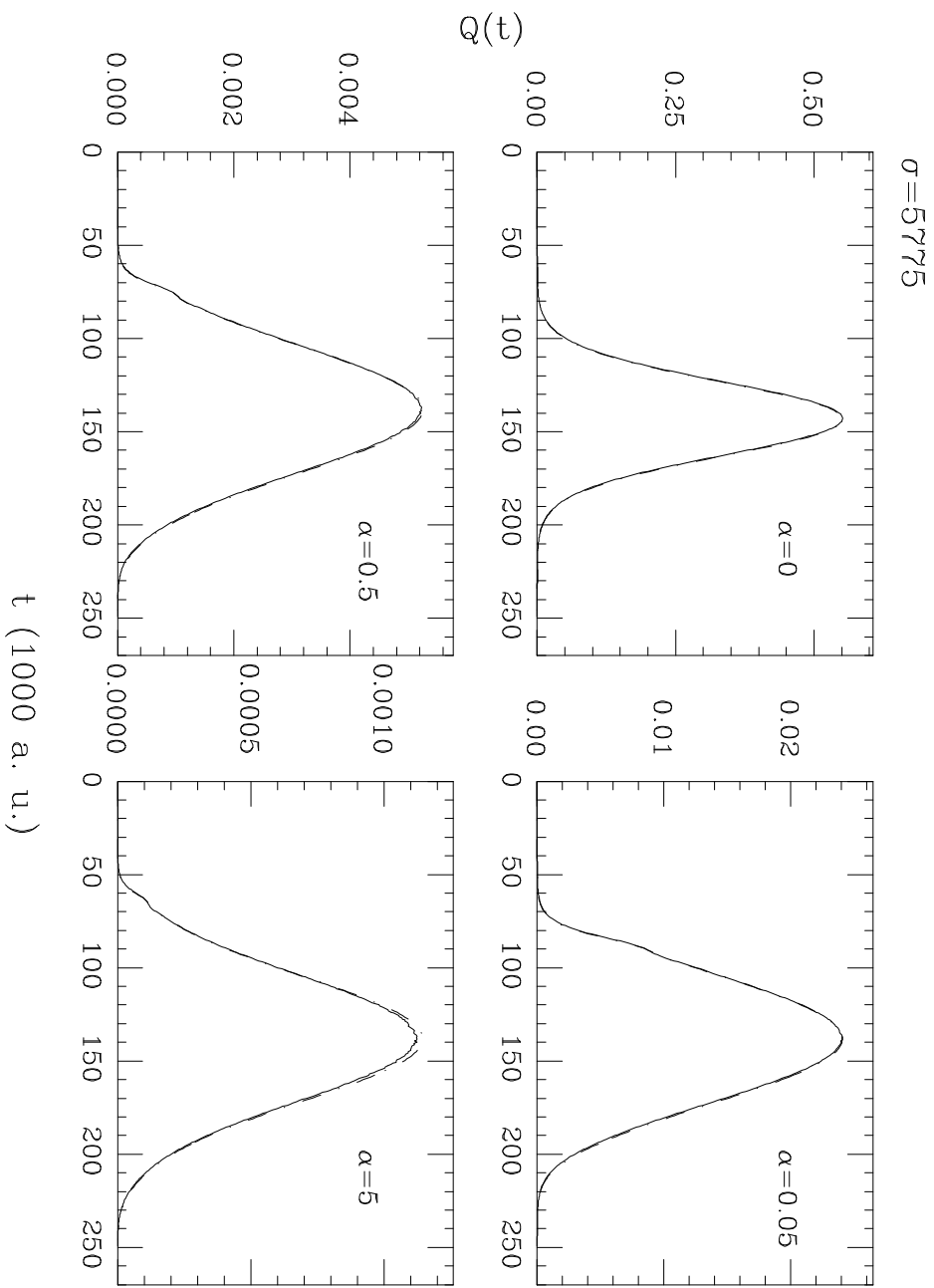

figure 1

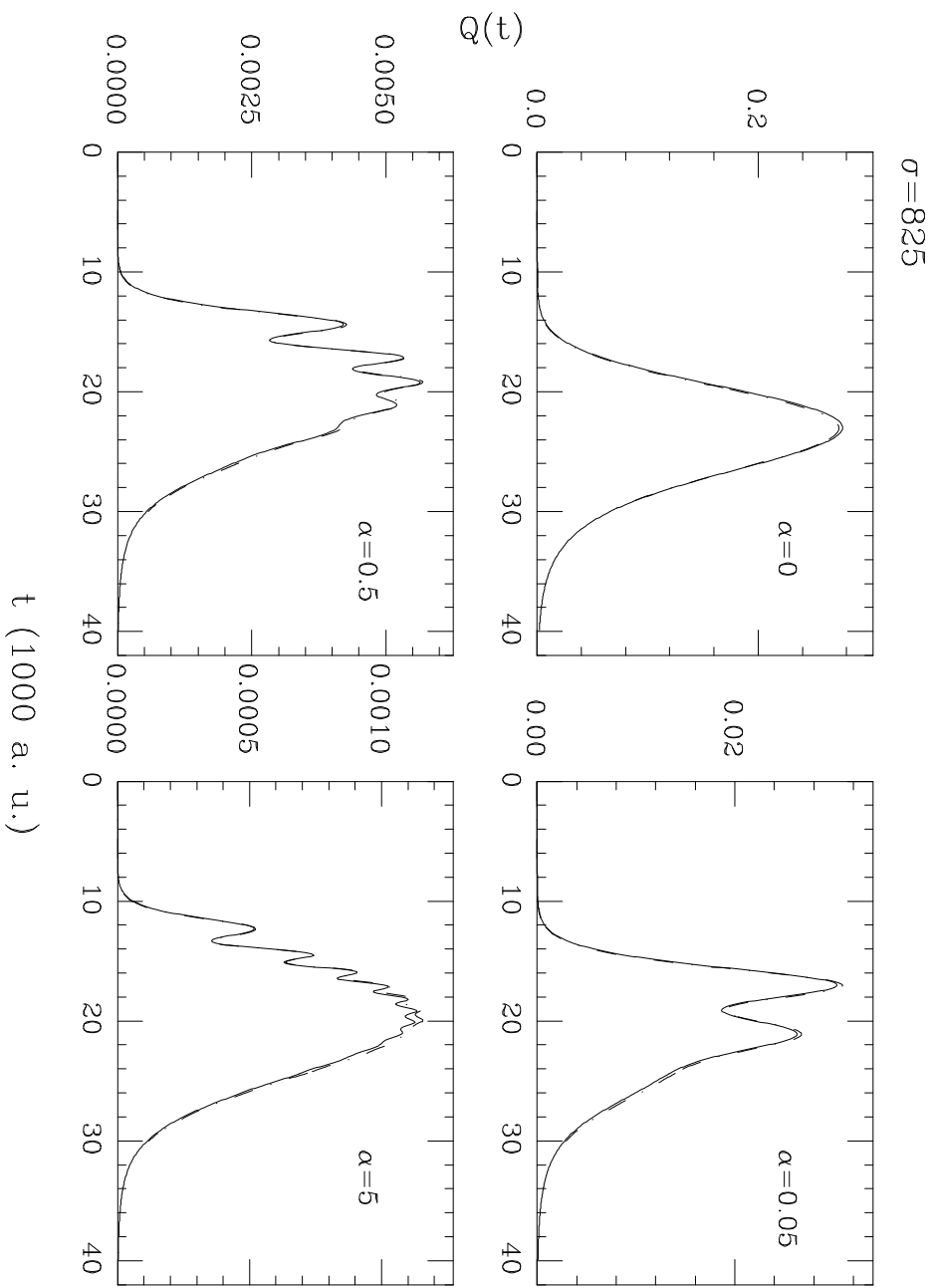

figure 2

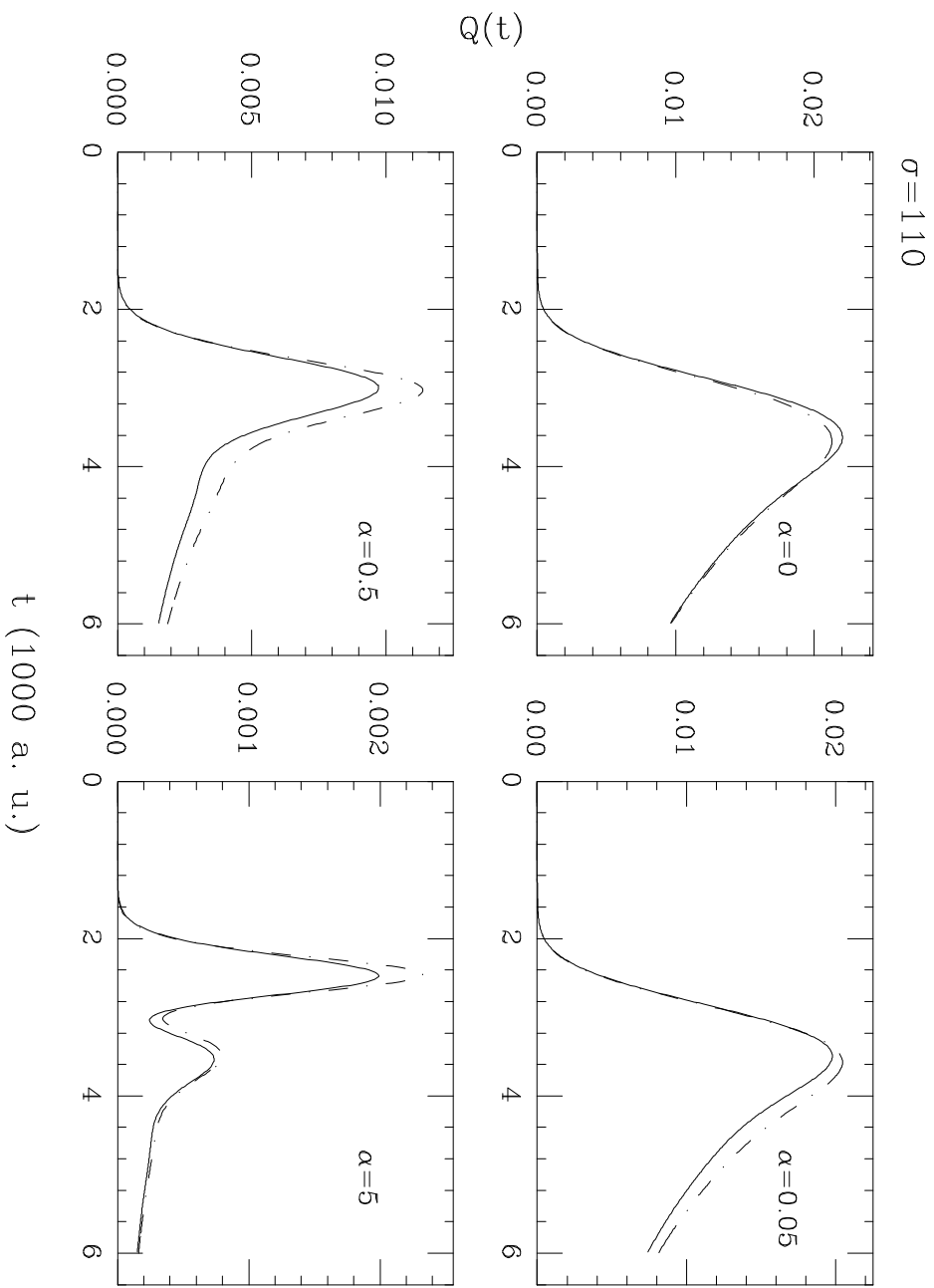

figure 3

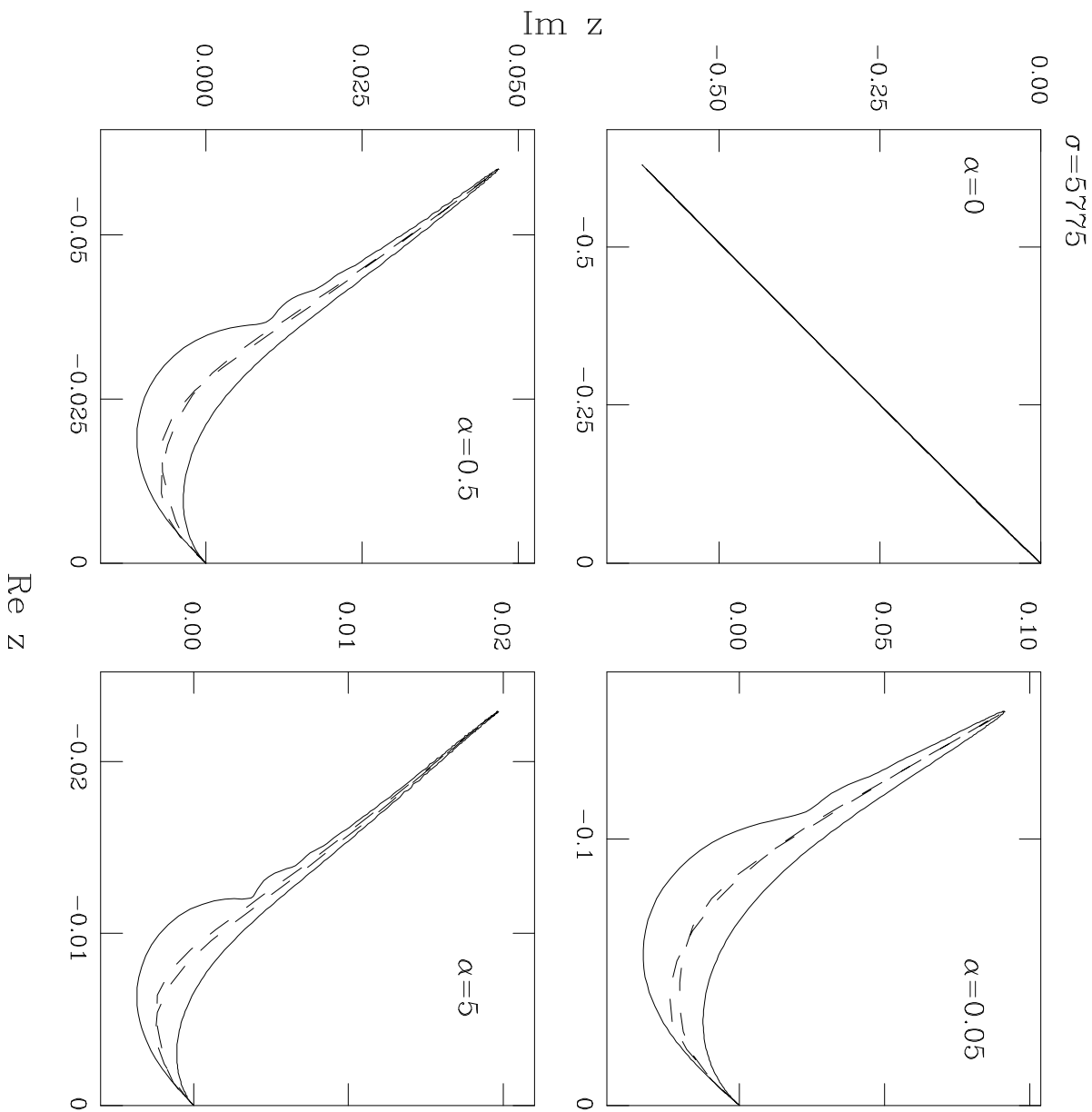

figure 4

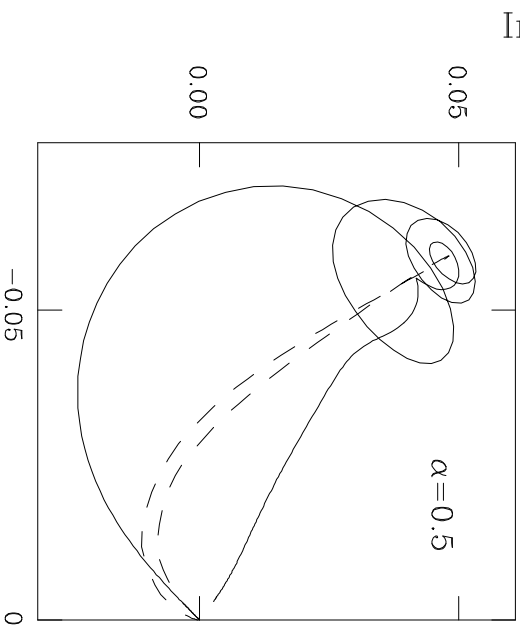
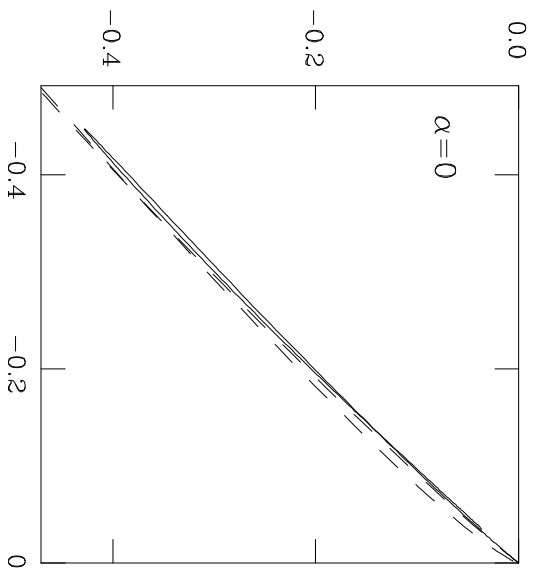
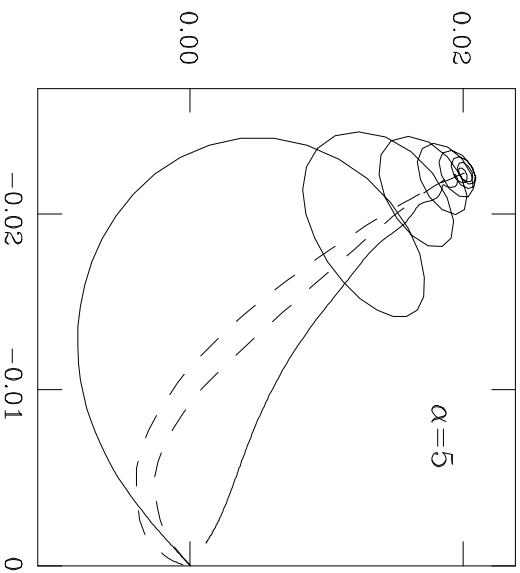
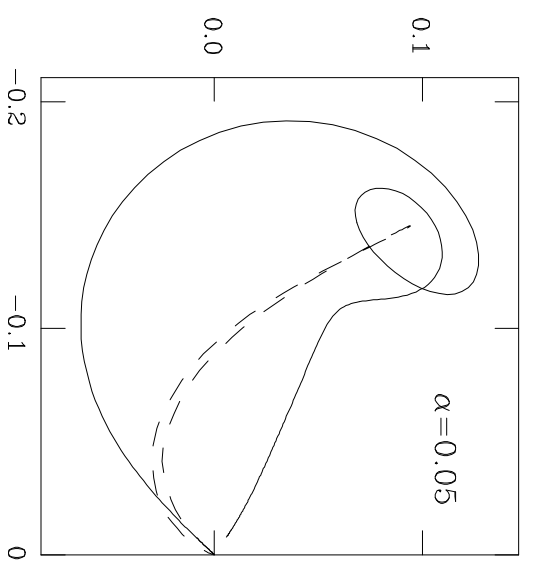

figure 5

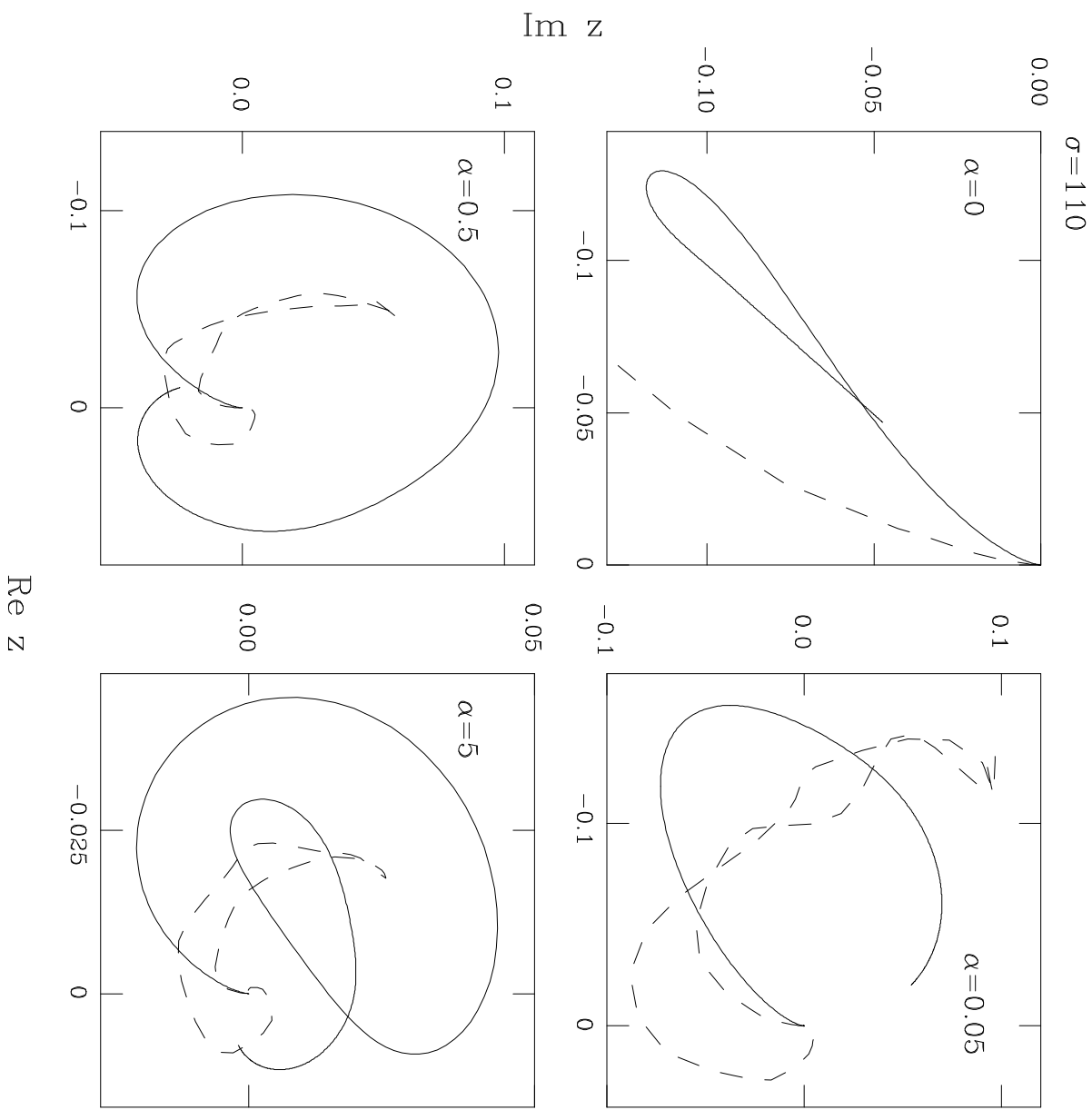

figure 6

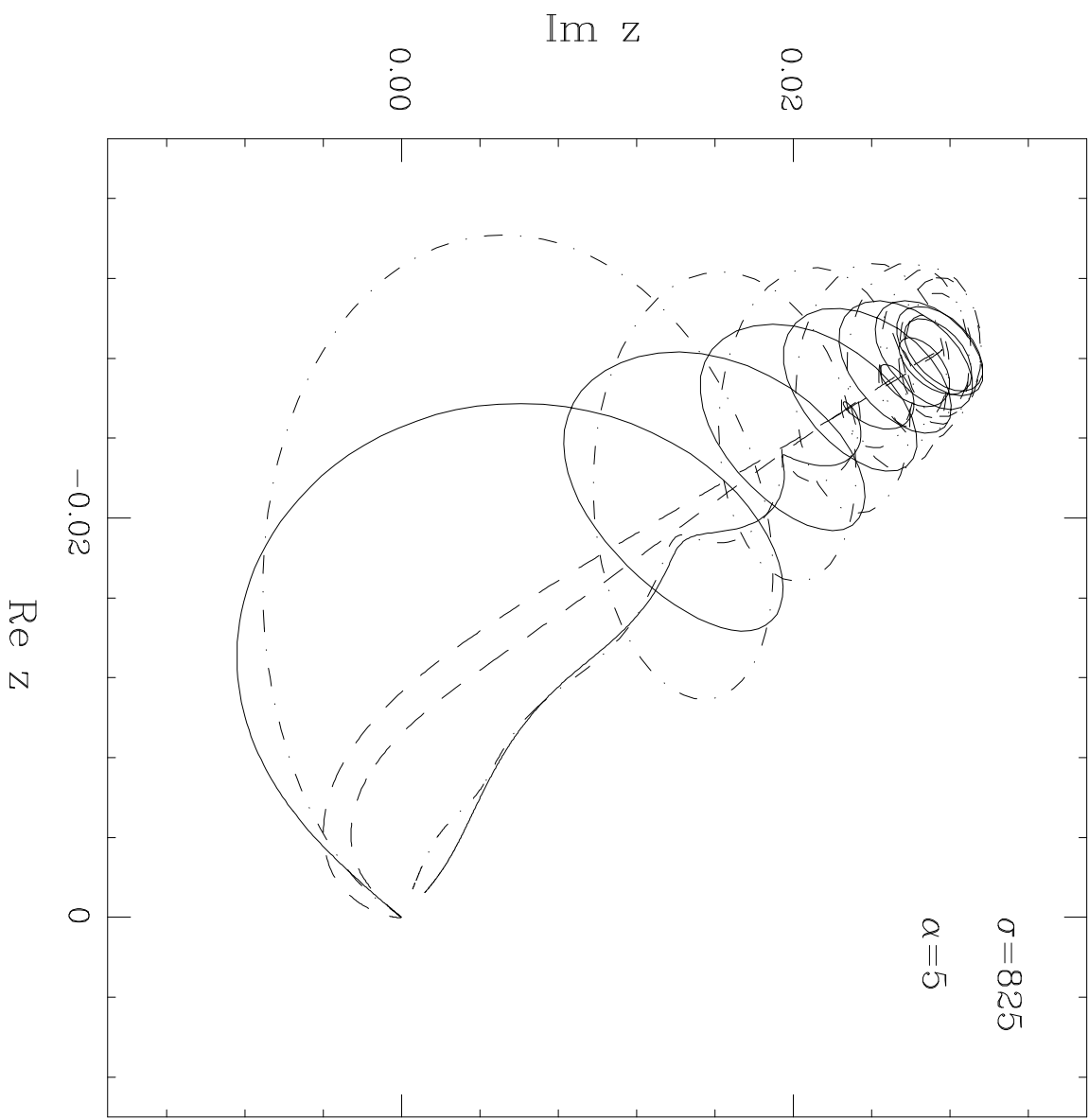

figure 7

$\sigma = 825$
$\alpha = 5$